\documentclass[pra,twocolumn,graphicx]{revtex4}
\usepackage{amssymb}
\usepackage{epsfig,amsmath}

\begin{document}

\title{Causality and quantum interference in time-delayed laser-induced nonsequential
double ionization}
\author{T. Shaaran$^1$, C. Figueira de Morisson Faria$^1$ and H. Schomerus$^2$}
\affiliation{$^1$Department of Physics and Astronomy, University College London, Gower
Street, London WC1E 6BT, United Kingdom\\$^2$Department of Physics,
Lancaster University,
Lancaster LA1 4YB,
United Kingdom }
\date{\today}
\begin{abstract}
We perform a detailed analysis of the importance of causality within the strong-field approximation and the steepest descent framework for the recollision-excitation with subsequent tunneling ionization (RESI) pathway in laser-induced nonsequential double ionization (NSDI). In this time-delayed pathway, an electron returns to its parent ion, and, by recolliding with the core, gives part of its kinetic energy to excite a second electron at a time $t^{\prime}$. The second electron then reaches the continuum at a later time $t$ by tunneling ionization. We show that, if $t^{\prime}$ and $t$ are complex, the condition that recollision of the first electron occurs before tunnel ionization of the second electron translates into boundary conditions for the steepest-descent contours, and thus puts constraints on the saddles to be taken when computing the RESI transition amplitudes. We also show that this generalized causality condition has a dramatic effect in the shapes of the RESI electron momentum distributions for few-cycle laser pulses. Physically, causality determines how the dominant sets of orbits  an electron returning to its parent ion can be combined with the dominant orbits of a second electron tunneling from an excited state. All features encountered are analyzed in terms of such orbits, and their quantum interference.
\end{abstract}

\maketitle

\section{Introduction}

Phenomena occurring in the interaction of matter with intense, low-frequency
laser fields, such as high-order harmonic generation (HHG), above-threshold
ionization (ATI), and laser-induced nonsequential double ionization (NSDI),
owe their existence to laser-induced recombination or rescattering processes,
in which a previously released electron interacts with its parent ion
\cite{Corkum}. The identification of their common dynamical origin has been
instrumental to exploit these effects in a number of applications, such as
for the generation of attosecond pulses \cite{atto1996} and the dynamic
imaging of matter with subfemtosecond precision \cite{Attoreview}. Since this
mechanism is intimately linked to specific dynamical pathways, it also led to
efficient semi-analytical approaches, such as the strong-field approximation
(SFA) (for seminal work on ionization and laser induced rescattering see, e.g., \cite{SFA} and \cite{SFAresc}, respectively). In this approach, the quantum-mechanical transition amplitude is
associated to the orbits of an electron returning to its parent ion. The
transition amplitude corresponding to a particular strong-field process takes
the form of a multiple integral, which, in many cases, can be solved
employing saddle-point methods \cite{SFApath}.

It is also a well-known fact that the orbits along which the
active electron returns typically occur in pairs. In particular,
thresholds can often be associated to conditions where the two
orbits of a given pair become almost degenerate. This violates the
saddle-point assumption which leads to expressions in terms of
individual orbits, but can be treated successfully in a uniform
approximation which describes the two orbits collectively
\cite{schomerussieber}. This approximation has been first applied in
strong-field physics to ATI \cite{Uniform2002} and, since then, has
been used in a wide range of phenomena, such as HHG
\cite{HHGuniform} and NSDI \cite{NSDI2004_1,NSDI2004_2,NSDI2005}.
This latter phenomenon is a typical example of a highly correlated
two-electron process occurring in strong laser fields. The physical
mechanism behind it is a three-step process in which the first
electron is released in the continuum by tunneling or multiphoton
ionization, and gains kinetic energy from the field. Subsequently,
it is driven back by the field towards its parent ion, with which it
rescatters. In this recollision, part of its kinetic energy is
transferred to the core, so that a second electron is freed.

There are several pathways through which NSDI may occur. Both electrons may,
for instance, be released simultaneously in a scattering process in which the
first electron, upon return, provides the second electron with enough energy
for it to overcome the second ionization potential. This process is known as
electron-impact ionization.
This pathway has been successfully used to explain a number of experimentally
observed features in the electron momentum distribution, such as peaks at
non-vanishing momenta \cite{VShape1,VShape2,NSDI2005}, and the recently
observed V-shaped structure which can been associated to the long-range
electron-electron repulsion \cite{NSDI2004_1,VShape5}.

 Apart from that,
the second electron may be released in a time delayed pathway, in
which the first electron promotes the second electron to an excited
bound state. Near the subsequent field maximum, the second electron
tunnels from this excited state. These pathways are becoming
increasingly important for two main reasons, which are directly
related to attosecond-imaging applications. First, the complexity of
studied NSDI targets is systematically increasing, and with this
internal excitations become more important. Second, delayed pathways
govern the below-threshold regime, for which the driving-field
intensities are too low for the second electron to leave by direct
ionization \cite{Belowthreshold}.

These time-delayed pathways are far less understood \cite{timedelayed}. Besides an overall
controversy relating to the physical mechanisms behind them (see, e.g.,
\cite{NSDIReview} for a detailed discussion of this topic), they can pose a
fundamental problem associated to causality, which we address in the present
work. A particularly relevant time-delayed pathway in NSDI is the
``recollision with subsequent tunneling ionization" (RESI). In previous work
\cite{SNF2010} it has been shown that this pathway can be understood as a
rescattered ATI-like process for the first electron, followed by direct ATI
for the second electron \footnote{In rescattered ATI, an electron collides
with its parent ion and loses part of its momentum before reaching the
detector, while in direct ATI an electron reaches the detector without
re-colliding. The momentum constraints encountered in both cases, however,
are very similar.}. RESI is a non-standard case within strong-field physics,
as the uniform approximation mentioned above requires some modifications. In
fact, a rigorous treatment of RESI within a saddle-point framework is a
non-trivial problem  as the rescattering of the first electron must precede
the ionization time of the second electron. So far, this issue has been
analyzed mainly in a classical framework, for which both times are real and
thus can be ordered. In the SFA, however, the associated orbits have complex
ionization times, which are required to account for the non-classical effect
of tunneling.

In this work, we address this complication of casuality for complex
 times  and investigate how it affects the momentum
distributions in the RESI  pathway of NSDI. We approach the problem
from the perspective of asymptotic expansions,  and resolve it by
explicit construction of steepest-descent integration contours. The
consequences of causality become apparent when one contrasts the
case of a purely monochromatic (and hence infinitely long) driving
field to the more realistic scenario of
few-cycle driving pulses.

This article is organized as follows. In Sec.~\ref{SFA} we briefly
recall the expression for the RESI transition amplitude and discuss the saddle point trajectories
with complex rescattering and ionization times. In Sec.~\ref{Momentum-space maps} we calculate the
individual momentum distributions of the first and the second
electron for monochromatic driving or a few-cycle pulse, disregarding the correlations imposed by causality.
In Sec.~\ref{Causality} we describe how
the causality requirement reflects itself in the complex time plane.
This description is then used in Sec.~\ref{Momentumdistributions} to
compute correlated two-electron momentum distributions.
Section \ref{Conclusions} contains our conclusions.

\section{Background}

\label{SFA}

\subsection{RESI transition amplitude}
The strong-field approximation (SFA) transition amplitude for RESI with final
electron momenta $\mathbf{p}_1$ and $\mathbf{p}_2$ reads \cite{Kopold2000,SF2010,SNF2010}.
\begin{eqnarray}\label{Mp}
&&M(\mathbf{p}_1,\mathbf{p}_2)  \\
&&\quad=\hspace*{-0.2cm}\iiint\limits_{t''<t'<t}\hspace*{-0.1cm}dt''\,dt'\,dt
\hspace*{-0.1cm}\int \hspace*{-0.1cm}d^{3}k
%\notag \\ &&{}\times
V_{\mathbf{p}_{2}}^{(e)}V_{\mathbf{p}_{1},\mathbf{k}}^{(eg)}V_{\mathbf{k}}^{(g)}
e^{iS(\mathbf{p}_1,\mathbf{p}_2,\mathbf{k},t'',t',t)},  \notag
\end{eqnarray}
with the action
\begin{eqnarray}
&&S(\mathbf{p}_1,\mathbf{p}_2,\mathbf{k},t'',t',t)
= \notag \\&&
 \quad
E_{\mathrm{1}}^{(g)}t''+E_{\mathrm{2}}^{(g)}t'+E_{\mathrm{2}}^{(e)}(t-t')-\int_{t''}^{t'}\hspace{-0.1cm}\frac{[%
\mathbf{k}+\mathbf{A}(\tau )]^{2}}{2}d\tau
 \notag \\&& \quad
-\int_{t'}^{\infty }\hspace{-0.1cm}\frac{[\mathbf{p}_{1}+%
\mathbf{A}(\tau )]^{2}}{2}d\tau
-\int_{t}^{\infty }\hspace{-0.1cm}\frac{[\mathbf{p}_{2}+\mathbf{A}(\tau
)]^{2}}{2}d\tau
  \label{singlecS}
\end{eqnarray}
and the form factors
\begin{eqnarray}
V_{\mathbf{k}}^{(g)}&=&\left\langle \mathbf{\tilde{k}}(t'')\right\vert V\left\vert \psi_1^{(g)}\right\rangle
 \notag \\ &=&\frac{1}{(2\pi
)^{3/2}}
\int d^{3}r_{1}V(\mathbf{r}_{1})e^{-i\mathbf{\tilde{k}}(t'')\cdot \mathbf{r}_{1}}\psi_1^{(g)}(\mathbf{r}_{1}),
\label{Vkg}
\end{eqnarray}
\begin{eqnarray}
V_{\mathbf{p}_{2}}^{(e)} &=&\left\langle \mathbf{\tilde{p}}_{2}\left( t\right)
\left\vert V_{\mathrm{ion}}\right\vert \psi _{2}^{(e)}\right\rangle
 \notag \\ &=&\frac{1}{(2\pi )^{3/2}}
\int d^{3}r_{2}V_{\mathrm{ion}}(\mathbf{r}_{2})e^{-i\mathbf{
\tilde{p}}_{2}(t)\cdot \mathbf{r}_{2}}\psi _{2}^{(e)}(\mathbf{r}_{2}),
\label{Vp2e}
\end{eqnarray}
and
\begin{eqnarray}
V_{\mathbf{p}_{1},\mathbf{k}}^{(eg)} &=&\left\langle \mathbf{\tilde{p}}_{1}\left(
t'\right) ,\psi _{2}^{(e)}\right\vert V_{12}\left\vert \mathbf{
\tilde{k}}(t'),\psi_{2}^{(g)}\right\rangle
 \notag \\  &=&\frac{1}{(2\pi )^{3}}
 \int \int d^{3}r_{2}d^{3}r_{1}\exp [-i(\mathbf{p}_{1}-\mathbf{k}
)\cdot \mathbf{r}_{1}]
 \notag \\ &&{}\times
 V_{12}(\mathbf{r}_{1,}\mathbf{r}_{2})[\psi_{2}^{(e)}(\mathbf{r}
_{2})]^{\ast }\psi_{2}^{(g)}(\mathbf{r}_{2}).\quad
  \label{Vp1e,kg}
\end{eqnarray}

Physically, Eq.\ (\ref{Mp}) is associated to a rescattering process in which
an electron, initially in a bound state $|\psi_1^{(g)}\rangle$ with energy
$E_1^{(g)}$, tunnels at time $t''$ into a Volkov state
$|\mathbf{\tilde{k}}(t^{\prime\prime})\rangle$. From the time $t''$ to the
time $t'$, this electron propagates in the continuum, until it is driven back
to its parent ion. Upon return, the electron scatters inelastically with the
core at time $t'$ and, through the interaction $V_{12}$, elevates the second
electron from the ground state $|\psi_2^{(g)}\rangle$ of the singly ionized
species (with energy $E_2^{(g)}$) to the excited state $|\psi_2^{(e)}\rangle$
(with energy $E_2^{(e)}$). Finally, at a later time $t$, the second electron
is released by tunneling ionization from the excited state
$|\psi_{e}^{(2)}\rangle$ into a Volkov state $|\mathbf{\tilde{p}}_{2}\left(
t\right)\rangle$. Here, $\mathbf{\tilde{k}}(t'')=\mathbf{k}$ and
$\mathbf{\tilde{p}}_2(t)=\mathbf{p}_2$ in the velocity gauge, and
$\mathbf{\tilde{k}}(t'')=\mathbf{k}+\mathbf{A}(t'')$,
$\mathbf{\tilde{p}}_2(t)=\mathbf{p}+\mathbf{A}(t)$ in the length gauge
\footnote{The length-to velocity gauge transformation will introduce a shift
$\mathbf{p}\rightarrow \mathbf{p}-\mathbf{A}(t)$, which will effectively
cancel out with the field dressing in the Volkov states for the
velocity-gauge situation. This issue will influence the ionization and
excitation prefactors, and is discussed in detail in \cite{SNF2010}.}. All
the information about the binding potential $V(\mathbf{r}_{1})$ of the first
electron and $V_{\mathrm{ion}}(\mathbf{r}_{2})$ of the second electron, and
the interaction $V_{12}(\mathbf{r}_{1},\mathbf{r}_{2})$ of the first electron
with the core, are embedded in the form factors (\ref{Vkg}), (\ref{Vp2e}) and
(\ref{Vp1e,kg}) respectively, which we will assume to be constant
over the parameter range in question.

Throughout this work, we will consider both a monochromatic, linearly polarized field,
for which
\begin{equation}
\mathbf{A}(t)=2\sqrt{U_{p}}\sin(\omega t)\hat{e}_{z}  \label{monochromatic}
\end{equation}
with ponderomotive potential $U_p$,
and a few-cycle pulse, for which
\begin{equation}
\mathbf{A}(t)=2\sqrt{U_{p}}\sin ^{2}\left( \frac{\omega t}{2N}\right) \sin
(\phi +\omega t)\hat{e}_{z},  \label{pulse}
\end{equation}
where $\hat{e}_{z}$ denotes the polarization vector, $N$ the number of cycles
in the pulse and $\phi $ the carrier-envelope phase. The associated electric
field is defined by $\mathbf{E}(t)=-\partial _{t}\mathbf{A}(t)$. A monochromatic wave is a reasonable approximation for long pulses \cite{NSDI2004_2}, or for few-cycle pulses when the carrier-envelope phase is integrated over.

\subsection{Saddle-point equations}
\label{saddlepointeqs}

For large driving-field intensities, Eq.\ (\ref{Mp}) is a strongly
oscillatory integral which can be evaluated using steepest-descent methods \cite{Bleistein}.
This requires to obtain the saddle points where the action  (\ref{singlecS})
is stationary,
\begin{equation}
\partial _{t''}S=\partial _{t'}S=\partial_{t}S=0, \quad\partial_{\mathbf{k}}S=\mathbf{0},
\end{equation}
%\begin{eqnarray}
%&&\partial _{t''}S(\mathbf{p}_1,\mathbf{p}_2,\mathbf{k},t'',t',t)=\partial %_{t'}S(\mathbf{p}_1,\mathbf{p}_2,\mathbf{k},t'',t',t)=0,
%\notag \\ &&
%\partial_{t}S(\mathbf{p}_1,\mathbf{p}_2,\mathbf{k},t'',t',t)=0, %\quad\partial_{\mathbf{k}}S(\mathbf{p}_1,\mathbf{p}_2,\mathbf{k},t'',t',t)=\mathbf{0},\notag \\ &&
%\end{eqnarray}
and establishes a direct link to semiclassical orbits with complex times and actions.
These orbits can be obtained efficiently by recognizing that the action splits into two independent parts,
\begin{eqnarray}
&&S_1(\mathbf{p}_{1},\mathbf{k},t'',t')
=E_{\mathrm{1}}^{(g)}t''+(E_{\mathrm{2}}^{(g)}-E_{\mathrm{2}}^{(e)})t'
  \notag \\ &&\quad
-\int_{t'}^{\infty }\hspace{-0.1cm}\frac{[\mathbf{p}_{1}+
\mathbf{A}(\tau )]^{2}}{2}d\tau
-\int_{t''}^{t'}\hspace{-0.1cm}\frac{[
\mathbf{k}+\mathbf{A}(\tau )]^{2}}{2}d\tau
 \label{Action1}
\end{eqnarray}
for the first electron
and
\begin{equation}
S_2(\mathbf{p}_{2},t)=-\int_{t}^{\infty }\hspace{-0.1cm}\frac{[\mathbf{p}
_{2}+\mathbf{A}(\tau )]^{2}}{2}d\tau +E_{\mathrm{2}}^{(e)}t  \label{Action2}
\end{equation}
for the second electron.
\subsubsection{First electron}

Explicitly, the stationary conditions upon $S_1$ lead to the equations
\begin{equation}
\left[ \mathbf{k}+\mathbf{A}(t'')\right] ^{2}=-2E_1^{(g)},  \label{saddle1}
\end{equation}
\begin{equation}
\mathbf{k=}-\frac{1}{t'-t''}\int_{t''}^{t'}d\tau \mathbf{A}(\tau ),  \label{saddle2}
\end{equation}
and
\begin{equation}
\lbrack \mathbf{p}_{1}+\mathbf{A}(t')]^{2}=\left[ \mathbf{k}+
\mathbf{A}(t')\right] ^{2}-2(E_2^{(g)}-E_2^{(e)}).
\label{saddle3}
\end{equation}

Condition (\ref{saddle1}) states the conservation law of energy for the first
electron when it reaches the continuum by tunneling. Condition
(\ref{saddle2}) constrains the intermediate momentum of the first electron so
that it returns to the site of its release, and also guarantees that the
intermediate momentum $\mathbf{k}$ of the electron is parallel to the laser
field. Condition (\ref{saddle3}) gives the conservation of energy upon
rescattering of the first electron, and states that the final kinetic energy
of the first electron is its kinetic energy upon return, minus the energy it
transferred to the core in order to excite the second electron. One should
note that, if $E_2^{(g)}=E_2^{(e)}$, the elastic rescattering condition for
high-order above-threshold ionization is recovered.

In particular, the solutions of the saddle point equations generally lead to
complex times, as Eq.~(\ref{saddle1}) admits no real solutions. This is a
consequence of the fact that tunneling is not a classically allowed process.
The imaginary part of $t''$ will be directly related to the width of the
barrier through which the electron tunnels:
the narrower the barrier, the smaller $\mathrm{Im}[t'']$ and the larger the tunneling probability.
Furthermore, for given values of momentum, rescattering may or may not have a
classical counterpart, depending on whether the maximal kinetic energy of the
first electron upon return is larger or smaller than the excitation energy
$E_{\mathrm{exc}}=E_2^{(g)}-E_2^{(e)}$. In order to analyze this aspect, it is
useful to recast Eq.~(\ref{saddle3}) in terms of the electron momentum
components $p_{1\parallel },p_{1\perp }$ parallel and perpendicular to the
laser-field polarization,
\begin{equation}
\lbrack p_{1\parallel }+A(t')]^{2}=\left[ \mathbf{k}+\mathbf{A}
(t')\right] ^{2}-2(E_2^{(g)}-E_2^{(e)})-p_{1\perp }^{2}.
\label{saddle3perp}
\end{equation}
This implies that a nonvanishing perpendicular momentum $p_{1\perp }$ shifts
the energy the first electron must provide to the core in order to excite the
second electron. The maximal kinetic energy upon return has to be larger than
an effective excitation energy
$\tilde{E}_{\mathrm{exc}}=(E_2^{(g)}-E_2^{(e)})+p_{1\perp }^{2}/2$ for the
rescattering process to have a classical counterpart. For that reason, one
expects that the classically allowed region in momentum space will be most
extensive for vanishing perpendicular momentum $p_{1\perp }$
\cite{SF2010,SNF2010}. As a function of the parallel component of momentum,
the most favorable rescattering conditions are then achieved when the first
electron leaves with $p_{1\parallel}=-A(t')$ around a maximum of $|A(t')|^2$,
of which there are two per field cycle, and returns near the subsequent field
crossing.

In terms of momentum constraints, this translates into the condition
$(p_{1\parallel},p_{1\perp})=(-A(t'),0)$ for the first electron around which the
partial momentum-space maps  in the $(p_{1\parallel},p_{1\perp})$ plane
discussed in Sec.~\ref{Momentum-space maps} will be centered.
For the first electron, the solutions of the saddle-point equations will
occur in pairs. These pairs correspond to the ``long" and the ``short" orbit
of an electron rescattering with its parent ion \cite{Saclay96}. Each cycle
will then contain two pairs of orbits, i.e., four orbits altogether. If a classically allowed region is
present, the rescattering times $t^{\prime}$ for each pair have vanishingly
small imaginary parts of opposite signs, i.e., they are located either in the
lower or upper complex half plane related to $t^{\prime}$. The start times
$t^{\prime\prime}$ will always exhibit nonvanishing and positive imaginary
parts. As one moves away from the center of a classically allowed region, the
saddles in a pair approach each other closely, until they almost coalesce at
its boundary.
If, on the other hand, the parameter range is such that no classically
allowed region is present, $\mathrm{Im}[t^{\prime}]$ will no longer be
vanishingly small in any momentum region. In this case, the physically
relevant saddle will be located in the upper half plane. The remaining saddle
will lead to exponentially increasing results and must be discarded (for a
detailed discussion see \cite{SF2010}).
Longer pairs of orbits, in which the
electron returns after having spent over a cycle in the continuum, will
contribute much less to the yield due to the spreading of the electronic wave
packet and therefore will be ignored.

\subsubsection{Second electron}
The stationarity condition upon $S_2(\mathbf{p}_{2},t)$ yields
\begin{equation}
\lbrack \mathbf{p}_{2}+\mathbf{A}(t)]^{2}=\mathbf{-}2E_2^{(e)},
\label{saddle4}
\end{equation}
which, physically, gives the energy conservation upon ionization of the
second electron.
This final ionization process is classically forbidden
throughout, as it always involves tunneling. If written in terms of the
parallel and perpendicular electron momentum components $p_{2\parallel
},\mathbf{p}_{2\perp },$ Eq. (\ref{saddle4}) reads
\begin{equation}
\lbrack p_{2\parallel }+A(t)]^{2}=\mathbf{-}2E_2^{(e)}-p_{2\perp }^{2}.
\label{saddle4perp}
\end{equation}
This implies that a non-vanishing transverse momentum $\mathbf{p}_{2\perp }$
effectively widens the potential barrier through which the second electron
must tunnel. A direct consequence will be an overall decrease in the yield
with increasing transverse momentum.   The electron tunnels most probably at
the field peak, and with $p_{2\perp}=0$. Hence, the momentum-space
conditions for which ionization of the second electron is most probable read
$(p_{2\parallel},p_{2\perp})=(0,0)$.  For the second electron, there exist two saddles per field cycle, which
do not coalesce. For vanishing momenta, the real parts of the corresponding
ionization times lie at subsequent maxima of the laser field, half a cycle
apart from each other. As the parallel electron momentum increases these
saddles move towards the field crossings. Since, however, their contributions
occupy the same region in momentum space, quantum mechanically they
interfere.

Equation (\ref{saddle4}) is identical to that
governing  direct ATI.
The second electron will therefore obey similar momentum constraints as
in this process (for which, however, causality does not play a role).
For instance, for a monochromatic field, momentum-resolved distributions should be
bounded by $\left\vert p_{2\parallel }\right\vert \leq 2\sqrt{U_{p}}$,
which corresponds to the traditional ATI cutoff energy of $2U_p$ \cite{SNF2010}.

\subsection{Causality and partial transition amplitudes}
The saddle point analysis described in the previous section provides two
independent sets of orbits, one set for the first electron, and another set
for the second electron. Using steepest descent methods, each set on its own
can be used to calculate separate momentum yields of the form
\begin{equation}
M^{(1)}(\mathbf{p}_{1})=\int_{-\infty }^{\infty}\hspace*{-0.3cm}dt''
\hspace*{-0.1cm}\int_{t''}^\infty\hspace*{-0.3cm}dt'\hspace*{-0.1cm}
\int d^{3}kV_{\mathbf{p}_{1},\mathbf{k}}^{(eg)}V_{
\mathbf{k}}^{(g)}e^{iS_1(\mathbf{p}_{1},\mathbf{k},t'',t')}  \label{Mp1}
\end{equation}
for the first electron, and
\begin{equation}
M^{(2)}(\mathbf{p}_{2})=\int_{-\infty}^{\infty }\hspace*{-0.2cm}dtV_{
\mathbf{p}_{2}}^{(e)}e^{iS_2(\mathbf{p}_{2},t)}  \label{Mp2}
\end{equation}
for the second electron (and we indeed do so in the following section).
However, the total RESI transition amplitude Eq.\ (\ref{Mp}) generally does not
factorize in this way,
\begin{equation}
M(\mathbf{p}_{1},\mathbf{p}_{2})\neq M^{(1)}(\mathbf{p}_{1})M^{(2)}(\mathbf{p}_{2}).
\end{equation}
The reason is the time constraint in the original expression (\ref{Mp}). In
this original integral, as well as in fully classical theories with
real-valued trajectories, the times $t''$, $t'$, and $t$ are all real and can
be ordered. However, the semiclassical pathways all have complex times.
Therefore, the underlying issue of causality requires closer attention. In
the remainder of this paper, we address this issue by examining the  key
technical step in the derivation of  semiclassical expressions, namely, the
construction of steepest descent contours in the complex-time plane. In order
to prepare this discussion, we set out in the following Sec.\ \ref{Momentum-space maps} by providing
semiclassical expressions for the partial amplitudes (\ref{Mp1}) and
(\ref{Mp2}).

\section{\label{Momentum-space maps}Partial momentum-space maps}

 In this section we provide a detailed analysis of
the partial one-electron transition probabilities
$|M^{(1)}(\mathbf{p}_{1})|^2$ and $|M^{(2)}(\mathbf{p}_{2})|^2$, as functions
of the momentum components $(p_{1,2\parallel},p_{1,2\perp})$. We evaluate the
partial probabilities asymptotically via the steepest-descent method, which
delivers expressions in terms of the saddles of the rapidly fluctuating
integrals in Eqs.\ (\ref{Mp1}) and (\ref{Mp2}), respectively. The
momentum-space maps obtained in this way will be useful for the construction
of the correlated RESI transition amplitude (\ref{Mp}),
Secs.\ \ref{Causality} and \ref{Momentumdistributions},
in which causality
must be taken into account.

For $M^{(2)}(\mathbf{p}_{2})$, the saddles are always well-separated in the momentum ranges of interest. Hence,
it suffices to employ the standard saddle-point approximation, in which the
saddles are treated individually and one performs a gaussian expansion around
each of them. For the partial transition amplitude $M^{(1)}(\mathbf{p}_{1})$
related to the first electron, however, the separation condition of saddles
is not always fulfilled. In fact, at the borders of the classically allowed
region in momentum space the saddles will approach each other  closely.
Hence, each pair of saddles must be treated collectively employing the
uniform approximation discussed in \cite{Uniform2002}. Beyond the boundary of
the classically allowed region, one of the two saddles in a pair will lead to
an exponentially increasing transition amplitude and thus must be left out.
This switching from two to one saddle near the boundary is known as a Stokes
transition \cite{Stokes}, and occurs via a bifurcation of the steepest
descent contour, which visits both saddles before the transition and only one
saddle after the transition. This happens when, for two saddles $a$ and $b$
defining a pair,
$\mathrm{Re}[S_1(\mathbf{p}_{1},\mathbf{k},t_{a}',t_{a}'')]=\mathrm{Re}[S_1(\mathbf{p}_{1},\mathbf{k}
,t_{b}',t_{b}'')]$. If, on the other hand, no classically allowed region
exists, one must take the saddle that leads to exponentially decaying
contributions throughout and the standard saddle-point approximation.
(Similar Stokes transition will be crucial for the analysis of causality in Sec. \ref{Causality}.)

\subsection{Monochromatic driving field}

\begin{figure}[t]
\begin{center}
\includegraphics[width=10cm]{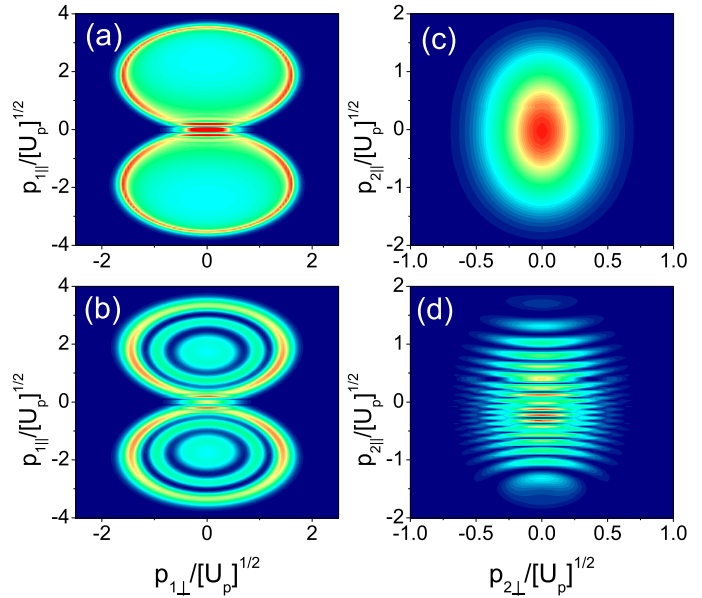}
\end{center}
\caption{Partial RESI transition probabilities for a monochromatic driving field,
as function of the momentum components $p_{1,2\parallel}$ parallel and
$p_{1,2\perp }$ perpendicular to the laser field
polarization. Panels (a) and (b) depict the
transition probability $|M^{(1)}(\mathbf{p}_{1})|^{2}$ for the first
electron, Eq.\ (\ref{Mp1}), while panels (c) and (d) give the transition probability
$|M^{(1)}(\mathbf{p} _{1})|^{2}$ for the second electron, Eq.\ (\ref{Mp2}).
Panel (a) shows the contribution
of the dominant trajectories of the first electron (one per half cycle), while panel (b) shows its interference
with the subdominant partner trajectories in the same half cycle, treated collectively in uniform approximation.
The trajectory pairs from the two half cycles result in identical distributions for
opposite parallel momenta; interference between these pairs of trajectories is negligible.
Panel (c) shows the contribution of an individual orbit of the second electron, while panel (d)
shows its interference with second orbit in the same field cycle. The distributions have been normalized with regard to the largest
value in each panel. Parameters are for
Helium ($ E^{(g)}_{1}=0.97$ a.u., $E^{(g)}_{2}=2$ a.u. and $E^{(e)}_{2}=0.5$ a.u.), and the
driving field is monochromatic with intensity $I=3\times
10^{14}\mathrm{W/cm^{2}}$ and frequency $ \protect\omega =0.057$ a.u.}  \label{Fig1}
\end{figure}

Figure \ref{Fig1} shows the resulting partial transition probabilities
$|M^{(1)}(\mathbf{p}_{1})|^2$ and $|M^{(2)}(\mathbf{p}_{2})|^2$ as functions
of the perpendicular and parallel momentum components of each electron, for a
monochromatic driving field.

We start with the simpler case of the second electron. For this electron,
there are two saddles, whose start times, for vanishing parallel momenta, lie
at adjacent field maxima separated by half a cycle of the driving field.
Each saddle on its own gives identical contributions to the probability
$|M^{(2)}( \mathbf{p}_{2})|^{2}$. These individual contributions are depicted
in Fig.~\ref{Fig1}(c). Figure~\ref{Fig1}(d) shows the rich interference
pattern which arises from the quantum mechanical superposition of the two
adjacent saddles. The individual contributions are maximal at
$(p_{2\parallel},p_{2\perp })=(0,0)$, for which the effective potential
barrier is narrowest and tunneling most probable. These momentum components
mark the center of the momentum map. The probability density quickly drops
off with increasing momenta $\mathbf{p}_2$, i.e., as one moves away from the
center of the momentum map. This is expected as the effective potential
barrier through which the electron tunnels  widens in this case (see
discussion of Eq.~(\ref{saddle4perp})).

For the first electron, there exist two pairs of saddles. Each pair stems
from a half cycle of the field and, for the parameter range employed in the
figure, can be associated to a classically allowed region centered at
$(p_{2\parallel },p_{2\perp })=(\mp 2\sqrt{U_p},0)$ (see discussion of
Eq.~(\ref{saddle3perp})). Figures \ref{Fig1}(a) and (b) show the resulting
probability $|M^{(1)}(\mathbf{p}_{1})|^{2}$, calculated either  only using the
saddle leading to exponentially decaying contributions outside the classical
boundary [panel (a)] or the complete pair [panel (b)]. The boundary of the
classical region is visible as an outer, bright ring, which indicates the
locus of the Stokes transition. If in panel (b) the second saddle would be
dropped abruptly beyond the transition one would obtain a cusp. We eliminated
this artifact by treating the pair collectively using the uniform
approximation \cite{Uniform2002}. The remaining fringes are caused by the
interference between the two orbits of the pair. In practice, there is only
interference between saddles separated by half a cycle around $(p_{1\parallel},p_{1\perp})=(0,0)$, as they mostly populate different momentum regions.

\subsection{\label{Momentum-space mapsPulse}Few-cycle pulse}

\begin{figure}[tbp]
\begin{center}
\includegraphics[width=8cm]{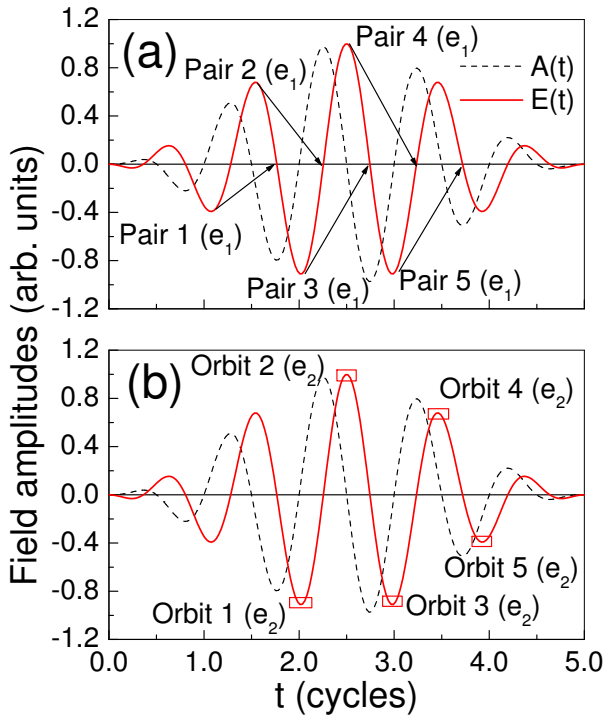}
\end{center}
\caption{Schematic representation of the electric field
$\mathbf{E}(t)$ and the corresponding vector potential
$\mathbf{A}(t)$, for a few-cycle pulse of five cycles. The arrows in
panel (a) indicate the approximate times around which the first
electron leaves, in case it returns at a crossing. The complex return and
start times for the indicated pairs of orbits will have real parts  in the
vicinity of such times. The rectangles in part (b) mark the
approximate ionization times for the second electron associated to
the orbits utilized in this work (Orbits 1 to 5). The fields have
been normalized to $E(t)/E_{0}$ and $A(t)/A_{0}$, where $E_{0}$,
$A_{0}$ denote the field amplitudes.} \label{Fig2}
\end{figure}

We will now turn to the few-cycle pulse given by Eq.~(\ref{pulse}), with
$N=5$ and carrier envelope phase $\phi=0$. This pulse is shown in
Fig.~\ref{Fig2}, together with the approximate ionization and rescattering
times for the first electron, and the ionization times for the second
electron [panels (a) and (b), respectively].

The first electron will leave close to an extremum of the field and return
near the subsequent field crossing [Fig.~\ref{Fig2}(a)]. The times indicated
in the figure are associated to the real parts of the solutions of the
saddle-point equations (\ref{saddle1})-(\ref{saddle4}), for $p_{1\perp}=0$
and $p_{1\parallel}=-A(t')$. Each of the five arrows is actually associated
to a pair of complex saddle points in the $(t'',t')$ plane. These pairs will
be referred to as Pairs $1(e_1)$ to $5(e_1)$.
Upon recollision, the returning electron will excite a second electron, which
will then leave near the subsequent field extremum. The orbits related to
these maxima are labeled Orbit $1(e_2)$ to $5(e_2)$. The
remaining orbits, which are not represented here, will play a negligible
role, due to the fact that the ionization times for the first or second
electron lie too close to the trailing edges of the pulse. This implies that
the corresponding ionization probabilities will be very small. Below, we
discuss momentum-space maps for the above-mentioned few-cycle driving pulse.
\begin{figure}[th]
\hspace*{-0.4cm}\includegraphics[width=9.8cm]{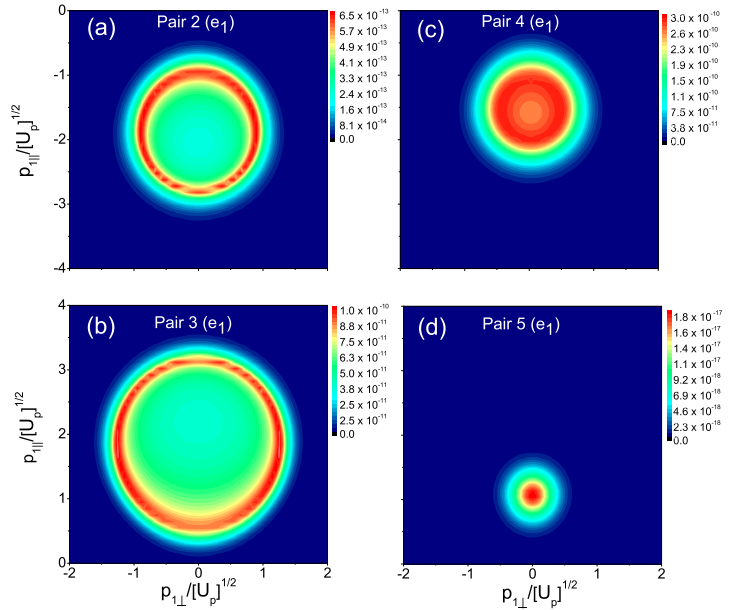}
\caption{
 Contributions from specific sets of orbits to the momentum-resolved RESI transition probabilities $|M^{(1)}(\mathbf{p}_{1})|^{2}$
of the first electron, Eq.\ (\ref{Mp1}), for the few-cycle pulse of Fig.\ \ref{Fig2}.
Panels
(a), (b), (c) and (d) correspond to Pairs 2, 3, 4 and 5$(e_1)$,
respectively (to eliminate distorting interference effects,
only the dominant orbit, which remains physical beyond the classical boundary, has been taken into account).
The pulse has
peak intensity $I=2.5\times 10^{14}\mathrm{W/cm^{2}}$, frequency $\omega
=0.057$ a.u. and carrier-envelope phase $\phi =0$; the other parameters are the same as in
Fig.\ \ref{Fig1}.
 }
\label{Fig3}
\end{figure}

We will start by analyzing the transition probability $|M^{(1)}(\mathbf{p}
_{1})|^{2}$ for the first electron as a function of $(p_{1\parallel
},p_{1\perp }),$ for each of the relevant pairs of orbits indicated in
Fig.~\ref{Fig2}. As in the monochromatic case, a reasonable insight into the
relevant momentum range is provided by the orbit in each pair whose
contribution decays exponentially outside the classically allowed region. In
Fig.~\ref{Fig3}, we display the contributions  of these orbits. As an overall feature, the
center of such maps is no longer located  at $(p_{1\parallel },p_{1\perp })=(\pm 2\sqrt{
U_{p}},0),$ but varies from cycle to cycle. This is due to the fact that this
estimate, albeit valid for a monochromatic field, no longer describes a field
crossing for a few-cycle pulse. In fact, the exact position of a field
crossing will depend on the pulse envelope and on the carrier-envelope phase.
Still, the momentum maps remain centered around vanishing transverse momenta
$p_{1\perp }.$ This is expected, as the effective widening of the excitation
energy for $p_{1\perp }\neq 0$ that can be inferred from
Eq.~(\ref{saddle3perp}) occurs regardless of the shape of the external
driving field.

Apart from that, the magnitudes of the partial probabilities
$|M^{(1)}(\mathbf{p}_{1})|^{2}$ will vary from cycle to cycle. Whether the
contributions of a certain pair will be prominent, irrelevant or even
vanishingly small will depend on several issues. If, for instance, the field
amplitude is large for a specific cycle, the ionization probability for the
first and second electrons are expected to be large as well. Thus, the
contributions from orbits starting at such times are expected to prevail.
Apart from that, if, for a particular set of orbits, the kinetic energy of
the electron upon return is much larger than the energy required to excite
the second electron, according to Eq.~(\ref{saddle3}) there will be an
extensive classically allowed momentum region. This will also play a role in
making the contributions of a particular set of orbits prominent. In the
specific case presented here, the contributions from Pair 3$(e_{1})$ and Pair
4$(e_{1})$, depicted in Fig.~\ref{Fig3}(b) and (c), respectively, are
comparable, and at least three orders of magnitude larger than those from the
other pairs. This is expected, as the ionization times related to both pairs
are very close to the center of the pulse (see Fig.~\ref{Fig2}). Hence, there
is a high tunneling probability for the electron at these times. Further
inspection, however, shows that the classically allowed region related to
Pair 3$(e_{1})$ is larger. This is related to the kinetic energy the electron
exhibits upon return, which is highest for this pair. The contributions of
the remaining pairs are less relevant, as the ionization times are closer to the trailing edge of the
pulse. For instance, the contributions from Pair
5$(e_{1})$, displayed in Fig.~\ref{Fig3}(d), are several orders of magnitude
smaller than the other pairs. This is due to the fact that for this specific
pair of orbits, the first electron does not return with enough kinetic energy
to excite the second electron and still reach the detector; hence,
rescattering is forbidden throughout. We have verified that this also happens
for Pair 1$(e_{1})$. For that reason, we do not include its contributions in
the present figure.
\begin{figure}[th]
\hspace*{-0.3cm}\includegraphics[width=9.2cm]{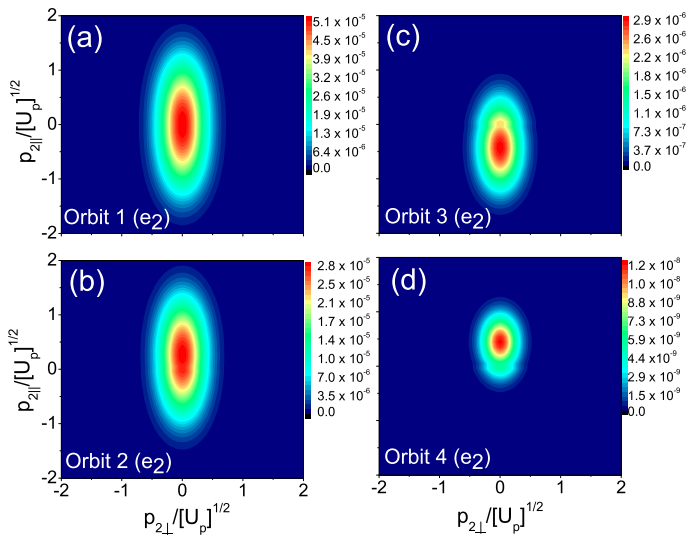}
\caption{Contributions of individual orbits to the momentum-resolved RESI transition probabilities
$|M^{(2)}(\mathbf{p}_{2})|^{2}$ for the second electron, Eq.~(\ref{Mp2}), for the same few cycle pulse
and atomic parameters as
in Fig.~\ref{Fig3}. Panels
(a), (b), (c) and (d) depict the contributions from
the orbits 1, 2, 3 and 4$(e_2)$, respectively (see Fig.~\ref{Fig2}).} \label{Fig4}
\end{figure}

We will now analyze the partial ionization probability of the second electron
by computing $|M^{(2)}(\mathbf{p}_{2})|^{2}$ for Orbits 1-4$(e_{2}) $
individually. Fig.~\ref{Fig4} shows the outcome of these computations as
functions of the parallel and perpendicular momenta $(p_{2\parallel
},p_{2\perp })$. Similarly to what has been observed for the first electron,
the centers of the momentum-space maps shift away from the position $(p_{2\parallel
},p_{2\perp })=(0,0)$. This shift is due to the lack of monochromaticity of the driving
field, which introduces an asymmetry around the field extrema and leads
to a slight bias towards either the positive or the negative parallel
momentum region. This bias is more pronounced for orbits whose ionization
times approach the trailing edge of the pulse. For the specific pulse
considered in this work, this can be observed in the contributions from
Orbits 3$(e_2)$ and 4$(e_2)$ [Fig.~\ref{Fig4}(c) and (d), respectively].
Still, for all cases, the momentum-space maps remain centered at $p_{2\perp
}=0.$ This is a consequence of the fact that, effectively, the potential
barrier through which the electron must tunnel is narrowest in this case,
regardless of the shape of the field (see discussion of
Eq.~(\ref{saddle4perp})).

Furthermore, the closer the ionization time is to the pulse center, the
larger the contributions from the corresponding orbit will be. For the
specific pulse considered here, for instance, the contributions from Orbit
1$(e_2)$ and Orbit 2$(e_2)$, displayed in Figs.~\ref{Fig4}(a) and (b),
respectively, are at least one order of magnitude larger than those of the
remaining orbits. In the figure, we do not include the contributions from
Orbit 5$(e_2)$, as these are at least four orders of magnitude smaller than
those of the remaining orbits.

\begin{figure}[th]
\begin{center}
\includegraphics[width=9cm]{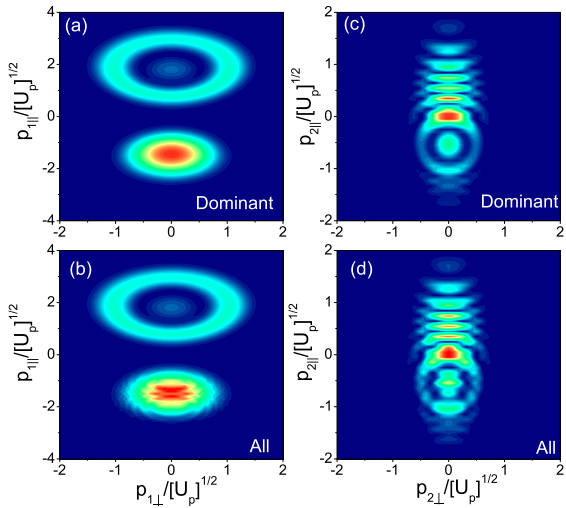}
\end{center}
\caption{Partial transition probabilities for a few-cycle pulse, as
function of the momentum components $p_{n\parallel}$ and $p_{n\perp }$.
Panels (a) and (b) depict the partial transition probability
$|M^{(1)}(\mathbf{p}_{1})|^{2}$ for the first electron,
while panels (c) and (d) depict $|M^{(2)}(\mathbf{p}_{2})|^{2}$
for the second electron. These are obtained by coherently adding the contributions
of orbits addressed in
Figs.~\ref{Fig3} and \ref{Fig4}
(the field and atomic parameters are the same as in these figures).
In panels (a) and (c), this sum is restricted to the dominant orbits of pairs
$3(e_1)$ and $4(e_1)$, as well as $1(e_2)$ and $2(e_2)$, respectively. The distributions have been normalized with regard to the maximum probability density in each panel.}
\label{Fig5}
\end{figure}

As in the monochromatic-field case, the contributions of the above-stated
sets of orbits must be added coherently. This will lead to interference
maxima and minima, both in the momentum maps and in the electron-momentum
distributions. In Fig.~\ref{Fig5}, we display the momentum maps for the
first and the second electron (left and right panels, respectively). The
upper and lower panels, respectively, exhibit the contributions of the
dominant sets of orbits, i.e., Pairs $3(e_1)$ and $4(e_1)$, and Orbits $1(e_2)$ and $2(e_2)$,
 and the overall partial momentum maps, in which all orbits have been included.

Figs.~\ref{Fig5}(a) and (b) exhibit two circular regions for which the
partial probability density is non-vanishing. These regions are roughly
centered at $(p_{1\parallel},p_{1\perp})=(\pm2\sqrt{U_p},0)$. Slight
displacements away from these points
are again related to the lack of monochromaticity of the field. The
negative parallel momentum region is dominated by the contributions of Pair
$4(e_1)$. In
fact, due to the large tunneling probability associated with it, it leads to
the brightest spot in these panels.
This pair interferes with Pair
$2(e_1)$, which is temporally displaced by a full cycle of the driving field;
this interference is visible as the substructure in Fig.~\ref{Fig5}(b). The
positive parallel momentum region is dominated by Pair $3(e_1)$. In both
panels, one may identify well-defined annular fringes,
which are caused by the interference between the long and the short orbit of
Pair $3(e_1)$. These orbits are temporally close, and their contributions are
several orders of magnitude larger than those of the remaining pairs  $1(e_1)$ and
$5(e_1)$.

The interference scenario is different for the second electron. In
this case, orbits located near different half cycles of the field lead to
contributions in overlapping momentum-space regions, roughly centered
at $(p_{2\parallel},p_{2\perp})=(0,0)$.
For instance, in Fig.~\ref{Fig5}(c), inclusion of
the dominant Orbits $1(e_2)$ and $2(e_2)$ already leads to a rich
interference pattern. Additional substructures appear if the remaining
orbits considered in Fig.~\ref{Fig4} are taken into consideration, as shown
in Fig.~\ref{Fig5}(d).

\section{\label{Causality}Causality for complex times}

In this section, we describe how the fact that ionization of the second
electron must be subsequent to the rescattering of the first electron
reflects itself in the complex time plane, and exemplify this for a
monochromatic driving field and a few-cycle pulse. This construction forms
the basis of the calculation of correlated RESI momentum distributions, which
are presented in Sec.~\ref{Momentumdistributions}.

Causality in the classical sense means that the first electron rescatters
with the core prior to the second electron being freed. If both the
rescattering time $t'$ of the first electron and the ionization time $t$ of
the second electron were real, this condition would simply require $t'<t$.
This condition is also embodied in the integration range of the SFA
expression (\ref{Mp}) for the transition amplitude. However, because
tunneling is not a classical process, the solutions of the saddle-point
equations ~(\ref{saddle1})-(\ref{saddle4}) are complex. Hence, in order to
evaluate the integrals in terms of these orbits one must determine how
causality manifests itself in the complex plane.

This can be done by reinspecting the steps that lead to the saddle point
approximation of the RESI transition amplitude (\ref{Mp}). This amplitude can
be rewritten as
\begin{eqnarray} \label{Mp1a}
&&\hspace*{-.5cm}M(\mathbf{p}_{1},\mathbf{p}_{2})
\\ &&  =\int_{-\infty }^{\infty}\hspace*{-0.3cm}dt''
\hspace*{-0.1cm}\int_{t''}^\infty\hspace*{-0.3cm}dt'\hspace*{-0.1cm}
\int d^{3}kV_{\mathbf{p}_{1},\mathbf{k}}^{(eg)}V_{
\mathbf{k}}^{(g)}e^{iS_1(\mathbf{p}_{1},\mathbf{k},t'',t')}
M^{(2)}(\mathbf{p}_{2};t'),  \notag
\end{eqnarray}
whereupon the time-conditioned amplitude
\begin{equation}
M^{(2)}(\mathbf{p}_{2};t')=\int_{t' }^{\infty }\hspace*{-0.2cm}dtV_{
\mathbf{p}_{2}}^{(e)}e^{iS_{2}(\mathbf{p}_{2},t)} \label{Mp2a}
\end{equation}
of the second electron depends on the rescattering time $t'$ via the lower
integration limit and is part of an integrand, in contrast to the partial
amplitude $M^{(2)}(\mathbf{p}_{2})=M^{(2)}(\mathbf{p}_{2};-\infty)$, Eq.\
(\ref{Mp2}), evaluated in Sec.\ \ref{Momentum-space maps}.

Let us assume that we have integrated out ${\bf k}$ via a saddle point
approximation (which is exact if one neglects the ${\bf k}$ dependence of the
form factors) and denote the resulting effective action as $\tilde
S(\mathbf{p}_1,\mathbf{p}_2,t'',t',t)$. We then can deform the integration
manifold over the three times $t''$, $t'$ and $t$ into the complex plane. The
integrand can be analytically continued and does not possess any poles. Therefore, the value of the integral does not change by this deformation. In
the steepest-descent method \cite{Bleistein}, this deformation is carried out
such that the quickly oscillating part $e^{i\tilde
S(\mathbf{p}_1,\mathbf{p}_2,t,t',t'')}$ of the integrand changes into a
smooth expression with maxima around the saddle points. To this end one
associates to each saddle $(t''_a,t'_a,t_a)$ a sheet on which $\mathrm{Re}\,\tilde
S(\mathbf{p}_1,\mathbf{p}_2,t,t',t'')=\mathrm{Re}\,\tilde
S(\mathbf{p}_1,\mathbf{p}_2,t_a,t'_a,t''_a)$. (In more than one dimension,
these sheets are not unique, but this does not affect any results
\cite{howls}.) Different sheets can meet in zeros of the integrand, since
there the phase is not well defined. Sheets have to be joined such that the
total deformed manifold starts and ends at the original lower and upper
integration limits of the integral, respectively. Therefore, one also needs
sheets connected to these boundaries, which are again constructed such that
the integrand decays rapidly as one moves away from the integration limits.
Following this prescription, not all saddle sheets will be part of the total deformed manifold,
which means that only a restricted set of all saddles is contributing to the
final expression. The saddle point approximation follows by expanding $\tilde
S$ around the maxima or integration limits, so that one obtains simple
integrals for each saddle point on the total contour, as well as each boundary
sheet. These individual contributions are accurate approximations if the
saddles and integration
limits are not too close to each other; otherwise one needs to
employ uniform approximations.

In order to apply this procedure to Eqs.\ (\ref{Mp1a}), (\ref{Mp2a}), we
proceed in three interlinked steps.

(I) Consider the time-conditioned amplitude $M^{(2)}(\mathbf{p}_{2};t')$. The
saddles in the complex $t$ plane are the same as for the partial amplitude
(\ref{Mp2}), and therefore determined by condition (\ref{saddle4}). However,
the steepest descent contour now depends on the integration limit $t'$, as we
have to construct a continuous contour in the complex $t$ plane that links it
to the upper integration limit (at real $+\infty$). The contour therefore
starts with a
 segment of fixed
$\mathrm{Re}[S_2(\mathbf{p}_{2},t)]=\mathrm{Re}[S_2(\mathbf{p}_{2},t')]$.
This boundary segment has then to be linked with contour segments passing
through saddles $t_a$, for which
$\mathrm{Re}[S_2(\mathbf{p}_{2},t)]=\mathrm{Re}[S_2(\mathbf{p}_{2},t_a)]$.
Compared to the calculation of the unconditioned amplitude
$M^{(2)}(\mathbf{p}_{2})$, this has two effects: (i) There is an additional
contribution from the boundary segment linked to $t'$, which we will neglect
as it is related to the electron-impact pathway \cite{NSDI2004_1}
and (ii) only a subset
of saddles $t_a$ contributing to $M^{(2)}(\mathbf{p}_{2})$ will lie on the
contour conditioned by starting at $t'$. Besides this selection criterion,
the individual saddle point contributions do not explicitly depend on $t'$.
However, the number of relevant saddles changes at Stokes transitions, which
for a given saddle $t_a$ occur when
$\mathrm{Re}[S_2(\mathbf{p}_{2},t')]=\mathrm{Re}[S_2(\mathbf{p}_{2},t_{a})]$.
(This condition is not sufficient; whether there is indeed a Stokes transition
can be verified by constructing the explicit integration contour, as carried
out below.)

(II) Next, we apply the saddle point approximation to the remaining integrals
over $t''$ and $t'$. We will focus at situations not too close to a Stokes
transition. Within a range in $t'$, the contributions from the saddle points
of $M^{(2)}(\mathbf{p}_{2};t')$ will then not depend on $t'$. Therefore, this
term can be treated as approximately constant, and will not affect the saddle
point conditions (\ref{saddle1})-(\ref{saddle3}) for $t''$ and $t'$\footnote{
However, the neglected contribution from the boundary segment will depend on
$t'$, with asymptotic dependence $\propto \exp[iS_2(\mathbf{p}_{2},t')]$. Its
saddles are therefore determined by
$S(\mathbf{p}_1,\mathbf{p}_2,\mathbf{k},t',t',t'')$. These saddles lie far
away from the real axis as the most favorable rescattering and ionization
conditions occur at different phases of the driving cycle (field crossing and
extremum, respectively). This suppresses the boundary contributions, but does
not affect the causality condition.}.

(III) This in turn means that we now can substitute the lower integration
limit in $M^{(2)}(\mathbf{p}_{2};t')$, which was considered general so far,
with the values of complex saddle points $t'_b$.

As an upshot, we find that for each saddle point trajectory ($t''_b,t'_b$) of
the first electron, only a certain number of saddle points $t_a$ of the
second electron will contribute. These are the ones that lie on the
continuous steepest descend contour in the complex $t$ plane that starts at
the rescattering time $t'_b$ of the first electron. Since the saddle-point
values $t'_b$  depend on the momentum $\mathbf{p}_1$ of the first electron,
but the remainder of the complex $t$ contour (and in particular the location
of the saddles $t_a$) depends on the  momentum $\mathbf{p}_2$ of the second
electron, this generalized causality requirement results in additional
correlations for the momentum distributions, which will be quantified in the
following Sec.\ \ref{Momentumdistributions}.
In the remainder of this section we focus on the explicit
construction of the described steepest descent contours.

\subsection{Steepest descent contours for monochromatic driving}

\begin{figure}[tbp]
\begin{center}
\includegraphics[width=7.5cm]{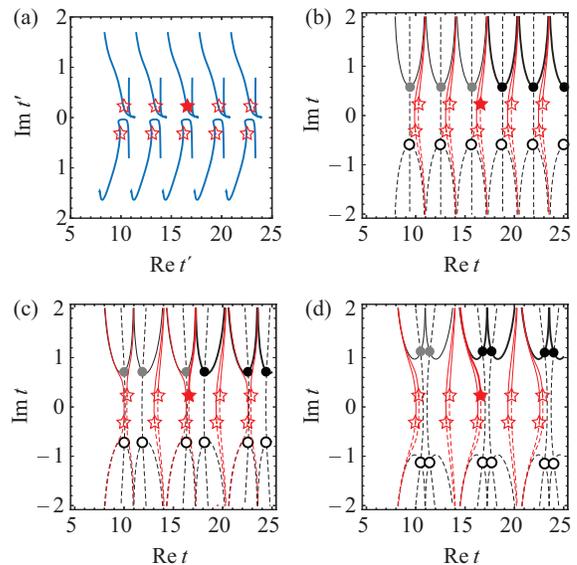}
\end{center}
\caption{Construction of steepest descent contours for the evaluation of
$M^{(2)}(\mathbf{p}_{2};t')$, Eq.\ (\ref{Mp2a}),
for monochromatic driving, with complex rescattering times $t'$ corresponding to
$p_{1\perp}=0$, $p_{1\parallel}=0$. In panel (a), these rescattering times,
obtained by solving the saddle point
equations (\ref{saddle1})-(\ref{saddle3}), are indicated by the star symbols.
For orientation, the curves show the trace of times $t'$ when $p_{1\parallel}$ is varied from $-5\sqrt{U_p}$
to $+5\sqrt{U_p}$ while $p_{1\perp}=0$ remains fixed.
In panels (b)-(d), the circles indicate the ionization times $t$ of the second electron for
$p_{2\perp}=0$, while the parallel momentum takes the values $p_{2\parallel}=0$ (panel b), $p_{2\parallel}=1.5\sqrt{U_p}$  (panel c),
and $p_{2\parallel}=3\sqrt{U_p}$  (panel d), respectively. These times are
obtained by solving the saddle point equation (\ref{saddle4}).
The solid circles indicate saddles that are physically allowed (with positive ${\rm Im}\,S_2$),
while the open circles are unphysical.
The dark solid lines indicate the steepest descent contour through these saddles, while the dark dashed lines
indicate steepest ascent contours, as well as steepest descent contour through unphysical saddles.
The red (light) lines are steepest descent contours passing through the rescattering times $t'$ (star symbols).
These lines are all determined from the condition ${\rm Re}\,S_2={\rm
const}$.
A full steepest descent contour must make use of solid lines in order to connect a lower integration limit
$t=t'$ (stars)
to the upper integration limit (at $t=+\infty$). The thick lines show the full contour for one selected
starting value, indicated by the solid star. This contour only visits a selection of all physical saddles (dark dots).
Between panels (c) and (d) a Stokes transition occurs, where an additional saddle becomes accessible.} \label{Figmono0}
\end{figure}

\begin{figure}[tbp]
\begin{center}
\includegraphics[width=7.5cm]{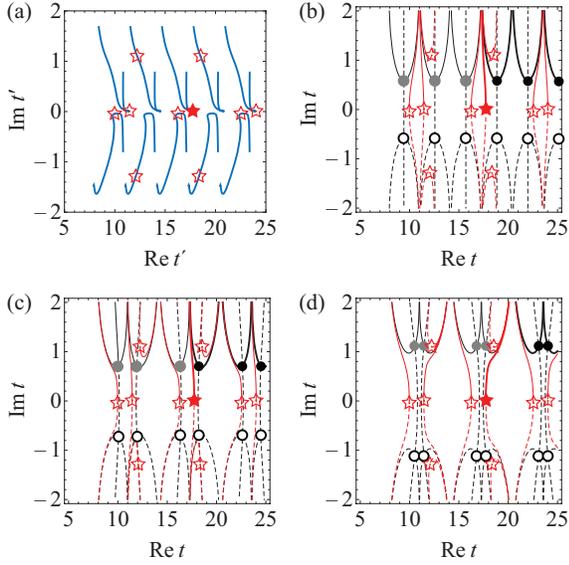}
\end{center}
\caption{Construction of steepest descent contours as in
Fig.\ \ref{Figmono0}, but for rescattering times corresponding to
$p_{1\perp}=0$, $p_{1\parallel}=2\sqrt{U_p}$. Due to this change,
in every cycle one pair of rescattering times
moves closer to the real axis, corresponding to physically allowed rescattering, while
the other pair moves away from the real axis.
The initial values of the steepest descent contours in panels (b)-(d) are
therefore shifted. For the rescattering time indicated by the solid star, again  a
Stokes transition occurs, but now a saddle is lost, not gained.}
\label{Figmono1}
\end{figure}

\begin{figure}[btp]
\begin{center}
\includegraphics[width=7.5cm]{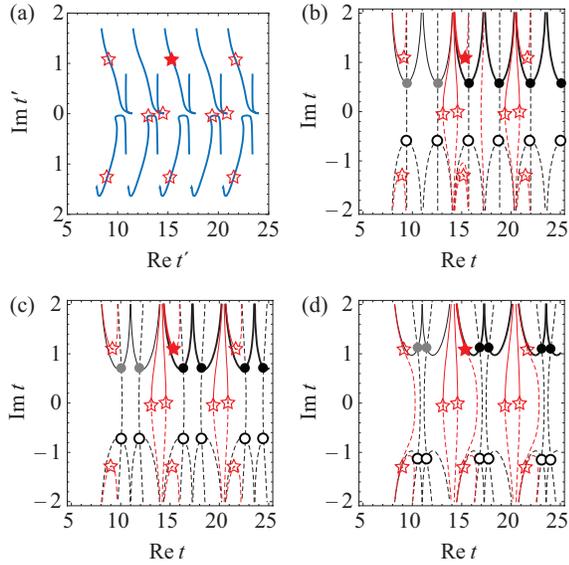}
\end{center}
\caption{Same as Fig.\ \ref{Figmono0}, but for rescattering times corresponding to
$p_{1\perp}=0$, $p_{1\parallel}=-2\sqrt{U_p}$. In comparison to Fig.\ \ref{Figmono1},
the rescattering times are shifted by a half cycle. The rescattering time indicated by the solid star
now corresponds to a negligible contribution
(its partner saddle is unphysical) and does not exhibit a Stokes transition.}
\label{Figmono2}
\end{figure}

Steepest descent contours for monochromatic driving and different
combinations of momenta are shown in Figs.\ \ref{Figmono0}-\ref{Figmono2}. In
order to identify the starting point of such contours, we show in panel (a)
of each figure the rescattering times $t'$ of the first electron (star
symbols), corresponding to $(p_{1\parallel},p_{1\perp})=(0,0)$ in Fig.\
\ref{Figmono0}, $(p_{1\parallel},p_{1\perp})=(2\sqrt{U_p},0)$ in Fig.\
\ref{Figmono1}, and $(p_{1\parallel},p_{1\perp})=(-2\sqrt{U_p},0)$ in Fig.\
\ref{Figmono2} (according to Fig.\ \ref{Fig1}, these values corresponds to
the point midway between the maxima in the momentum map, as well as the locations
of these maxima). Panels (b)-(d) of Figs.\ \ref{Figmono0}-\ref{Figmono2} show
these values as integration limits in the $t$ plane, along with the complex
ionization times of the second electrons (dots).  In all panels
$p_{2\perp}=0$, while the parallel momentum takes the values
$p_{2\parallel}=0$ (panel b), $p_{2\parallel}=1.5\sqrt{U_p}$ (panel c), and
$p_{2\parallel}=3\sqrt{U_p}$ (panel d), respectively [according to Fig.\
\ref{Fig1}, panel (b) therefore corresponds to the maximum in the momentum
map of the second electron]. The solid lines in these plots show steepest
descent segments from which a full contour must be constructed. These only
pass saddles in the upper half plane (solid dots), as the saddles in the
lower half plane (open dots) have ${\rm Im}\, S_2<0$. They will lead to exponentially increasing contributions and therefore are
unphysical. However, for a given starting value not all the physical saddles
are visited. This is illustrated by  sample contours (thick lines) that start
at a selected rescattering time, indicated by the solid star, and only pass
the saddles indicated by the dark dots, while the gray solid dots are left
out.

In this way, for any given saddle-point rescattering time $t'$ we can define
a clear boundary in the complex $t$ plane that determines whether a
saddle-point ionization time is allowed by causality or not. This boundary
passes over saddles whenever a Stokes transition occurs. The thick sample
contours display a Stokes transition in Figs.\ \ref{Figmono0} and
\ref{Figmono1}, but not in \ref{Figmono2}.

Inspecting these figures more
generally, we see that to a good approximation, the causality requirement
assumes the form ${\rm Re}\,t'<{\rm Re}\,t$. This is the case because Stokes
transitions take place when a start time $t'$ crosses over the
steepest-\emph{ascent} segment of a physically allowed ionization time $t$.
In the figures, these lines are seen to lead almost vertically from a
physical saddle in the upper half of the complex-$t$ plane to an unphysical
mirror saddle in the lower half plane. We verified that this approximate
causality criterion remains valid over the range of momenta where the
momentum maps in Fig.\ \ref{Fig1} are large.

\subsection{Steepest descent contours for a few-cycle pulse}

\begin{figure}[tbp]
\begin{center}
\includegraphics[width=8cm]{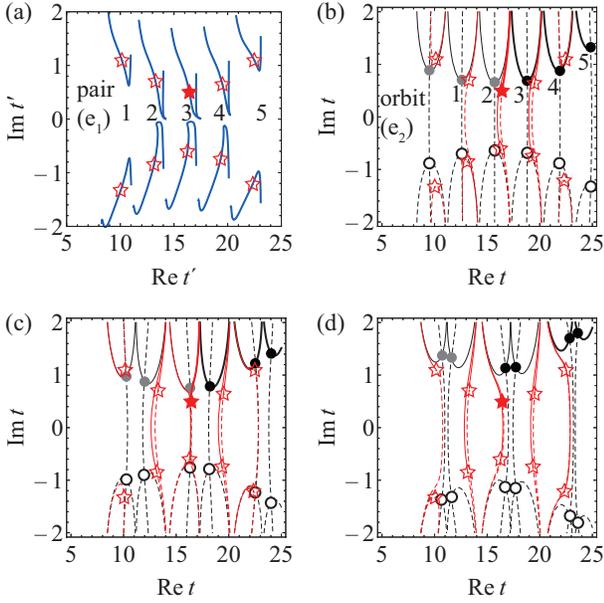}
\end{center}
\caption{Construction of steepest descent contours as in Fig.\ \ref{Figmono0}, but for a few cycle pulse.
Labels refer to the orbit pairs (first electron) and orbits (second electron) indicated in Fig.\ \ref{Fig2}.
Panel (a) shows the effect of the finite pulse length on the first electron.
The stars again denote the rescattering times for $p_{1\perp}=0$,
$p_{1\parallel}=0$. In the tails of the pulse the driving is weak, resulting in rescattering
times that are shifted away from the real axis.
Therefore, some of these trajectories of the first electron become
negligible. Panels (b)-(d) show that the same effect also occurs for the ionization times $t$ of the second electron.
The solid star indicates a rescattering time in the center of the pulse, which
gives the largest semiclassical contribution.
The value of this time is close to that in the monochromatic case, and
the associated steepest descent contours (thick lines) display
the same type of Stokes transition as in that case (cf.\ Fig.\ \ref{Figmono0}).
} \label{Figpulse0}
\end{figure}

\begin{figure}[tbp]
\begin{center}
\includegraphics[width=8cm]{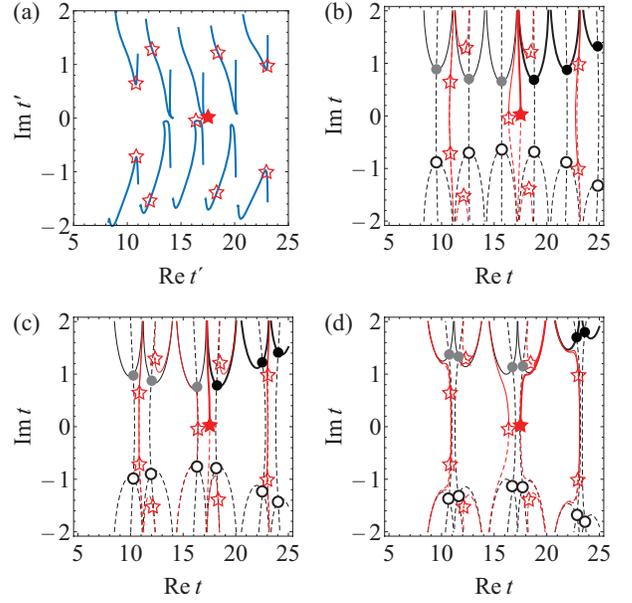}
\end{center}
\caption{Construction of steepest descent contours for a few cycle pulse
as in Fig.\ \ref{Figpulse0}, but corresponding to
$p_{1\perp}=0$, $p_{1\parallel}=2\sqrt{U_p}$, for which the
rescattering time indicated by the solid star moves closer to the real axis.
In comparison to the monochromatic case in Fig.\ \ref{Figmono1}, the associated steepest descent contours
again display the same type of Stokes transition.
} \label{Figpulse1}
\end{figure}

\begin{figure}[tbp]
\begin{center}
\includegraphics[width=8cm]{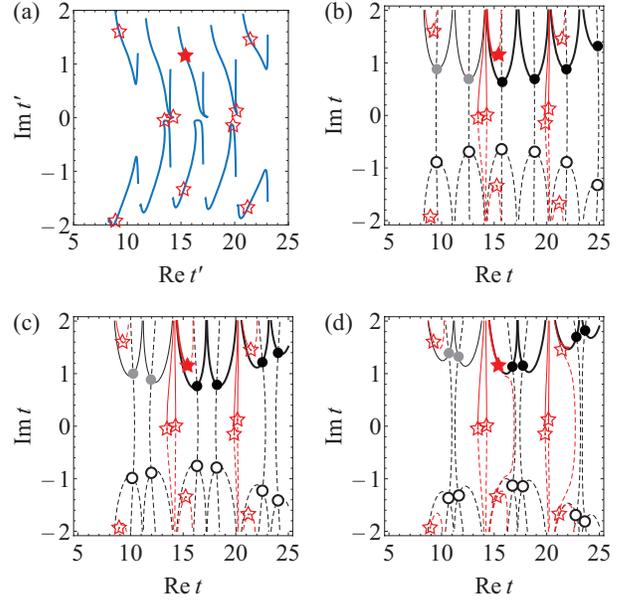}
\end{center}
\caption{Same as Fig.\ \ref{Figpulse0}, but for rescattering times corresponding to
$p_{1\perp}=0$, $p_{1\parallel}=-2\sqrt{U_p}$. As in the monochromatic case (Fig.\ \ref{Figmono2}),
the rescattering time indicated by the solid star
now corresponds to a negligible contribution and does not exhibit a Stokes transition.
} \label{Figpulse2}
\end{figure}

Figures \ref{Figpulse0}-\ref{Figpulse2} show the corresponding steepest
descent contours for the few cycle pulse. Instead of a periodic repetition
cycle-by-cycle, the rescattering times (stars) and ionization times (dots)
are now modulated according to their position in the pulse (see Sec.\
\ref{Momentum-space mapsPulse} and Fig.\ \ref{Fig2}). In the tails of the
pulse, where the intensity is small, the rescattering and ionization times
move away from the real axis. Compared to the monochromatic case, the
steepest descent contours are therefore distorted, but the main features
are still strikingly similar. In particular, we observe the same type of
Strokes transitions, and the simplified causality criterion ${\rm
Re}\,t'<{\rm Re}\,t$ is clearly still a good approximation.  However, from
these figures we can still anticipate that the causality constraint in itself
plays a much more significant role when it comes to the calculation of actual
transition probabilities. Large contributions to the momentum yield should
stem from the saddle points close to the center of the pulse. Because of the
causality constraint, however, it is not always possible to combine the most favorable
rescattering and ionization times. In the figures, this is illustrated by the
sample contours, which start at the rescattering time $t'$ of the long orbit
in pair $4(e_1)$ (the most favorable time according to Fig.\ \ref{Fig3}). The
most favorable ionization time for the second electron is that of orbit
$2(e_2)$ (see Fig.\ \ref{Fig4}), but these orbits generally cannot be
combined  because of causality. In the following section, we quantify
the consequences in terms of the correlated momentum distributions of both electrons.

\section{\label{Momentumdistributions}Momentum distributions}

In this section we compute two-electron momentum distributions as functions
of the momentum components $(p_{1\parallel},p_{2\parallel})$ parallel to the
laser field polarization, contrasting again the case
of a monochromatic field to that of a few cycle pulse.
For both cases, we calculate the momentum distribution
for resolved parallel and restricted ranges of the transverse momenta. If
transverse momenta are fully integrated out, as it is done in many NSDI
studies, quantum-interference effects get washed out and causality-related
effects can no longer be identified. This also holds for cusps or further
artifacts that may be present in saddle-point approximations, and which
indicate their breakdown. Hence, transverse momentum integration would mask
the very effects we intend to analyze.

We consider the distributions
\begin{equation}
F(p_{1\parallel},p_{2\parallel})=\int^{p^{(\mathrm{max})}_{1\perp}}_{p^{(\mathrm{min})}_{1\perp}}
\hspace*{-0.4cm}\int^{p^{(\mathrm{max})}_{2\perp}}_{p^{(\mathrm{min})}_{2\perp}}
\hspace*{-0.5cm}d^{2}p_{1\perp}d^{2}p_{2\perp}|M(\mathbf{p_1},\mathbf{p_2})+\mathbf{p_1}\leftrightarrow \mathbf{p_2}|^2,
\end{equation}
which have been symmetrized with regard to electron exchange. This is necessary as the two electrons are indistinguishable.
In the above-stated equation, $p^{(\mathrm{min})}_{n\perp}$ and
$p^{(\mathrm{max})}_{n\perp}$ $(n=1,2)$ denote the minimal and the
maximal transverse momenta to be taken into account.

\begin{figure*}
\noindent\hspace*{-0.5cm}\includegraphics[width=15cm]{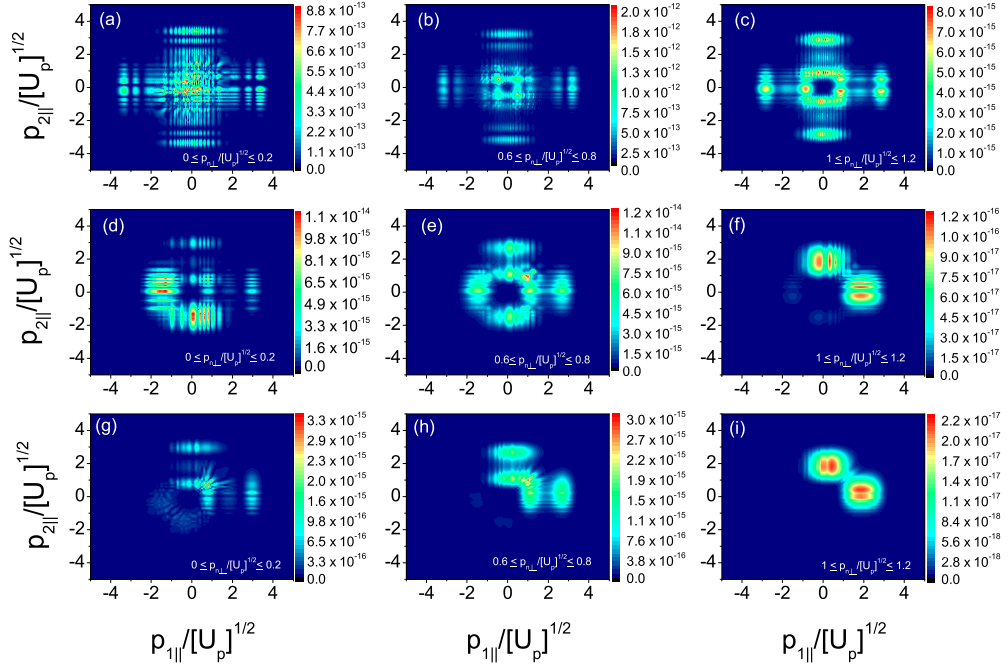} \caption{
Correlated RESI two-electron momentum distributions
$F(p_{1\parallel},p_{2\parallel})$ as functions of the
electron momentum components $p_{n\parallel}$ ($n=1,2$) parallel to the
laser-field polarization. The atomic parameters correspond to Helium (see
Fig.\ \ref{Fig1}). Panels (a) to (c) have been computed for a monochromatic field of intensity $I=3 \times 10^{14}\mathrm{W/cm}^2$ and frequency $\omega=0.057$ a.u., while in panels (d) to (i) a five-cycle pulse of intensity $I=2.5 \times 10^{14}\mathrm{W/cm}^2$,  frequency $\omega=0.057$ a.u. and carrier-envelope phase $\phi=0$ has been used.
The middle row of
panels (d)-(f) show results for the few-cycle pulse if causality is discarded
(these results are therefore unphysical), while the bottom panels (g)-(i)
show the results for the few-cycle pulse including the causality restrictions. From the left to the
right column of panels, we integrated over a transverse momentum range
centered around low, medium, and large transverse momenta, respectively.}
\label{Fig7}
\end{figure*}

For a monochromatic driving field, the causality-induced shielding of saddles
does not have an observable effect on the momentum distributions. This is so because the saddles are repeating  periodically cycle by cycle, and all have to be added coherently,. This eventually
results in a closely spaced train of delta functions which embody the Bohr
resonance condition for multiple-photon absorption \cite{kopold}. Smoothing
over these delta functions, we find
$M(\mathbf{p}_1,\mathbf{p}_2)=M^{(1)}(\mathbf{p}_1)M^{(2)}(\mathbf{p}_2)$ as
if causality had been ignored. For a few-cycle pulse, however, the orbits in
different cycles are not equivalent, and casuality has observable
consequence. This is demonstrated in Fig.~\ref{Fig7}, where we present the
distributions $F(p_{1\parallel},p_{2\parallel})$
for a monochromatic field (upper panels) and a few-cycle pulse
(middle panels ignoring causality, and lower panels respecting casuality),
for increasing transverse momenta (left panels: `low momenta'
$0<p_{1\perp},p_{2\perp}<0.2\sqrt{U_p}$; middle panels: `medium momenta'
$0.6\sqrt{U_p}<p_{1\perp},p_{2\perp}<0.8\sqrt{U_p}$; right panels panels:
`large momenta' $\sqrt{U_p}<p_{1\perp},p_{2\perp}<1.2\sqrt{U_p}$).

For the monochromatic driving field, Figs.~\ref{Fig7}(a)-(c), all
distributions adhere to the momentum constraints defined
for RESI \cite{SF2010}, i.e., they are cross-shaped and located along the axes $p_{n\parallel}=0$ $(n=1,2)$.  The elongation of such distributions is determined by the saddle-point equation (\ref{saddle3perp}), which yields the momentum-space regions filled by the rescattering of the first electron, and the width of such distributions is given by the maximal and minimal momenta determined by Eq.~(\ref{saddle4perp}).  They are also symmetric upon the reflection
$p_{n\parallel}\rightarrow -p_{n\parallel}$, which is a direct consequence of
the time-reversal symmetry of the field. This means that, without symmetrization upon electron exchange, these distributions would be located along the horizontal axis $p_{1\parallel}=0$. The symmetrization  $\mathbf{p}_1 \leftrightarrow \mathbf{p}_2$ leads to the vertical axis of the crosses.  The rather rich substructure present
in the figure is associated to the quantum interference between the long and
short orbit for the first electron, and between orbits displaced by half a
cycle for the second electron.

%%%%%%%%%%%%%%%%%%%%%%%%%%%%%%%%%%%%%%%%%%%%%%%%%%%%%%%%%%%%%%%%%%%%%5555555
For a few-cycle pulse, in contrast, the electron momentum distributions are
very much affected by causality. To isolate these effects, we start by
analyzing the situation if causality could be neglected, displayed in
Figs.~\ref{Fig7}(d), (e) and (f). These distributions are again obtained by
assuming
$M(\mathbf{p}_1,\mathbf{p}_2)=M^{(1)}(\mathbf{p}_1)M^{(2)}(\mathbf{p}_2)$.
The distributions are then asymmetric, which is a direct consequence of the
asymmetries in the momentum maps in Fig.~\ref{Fig5}.
The momentum maps are localized in the momentum region determined by the
dominant orbits, located close to the center of the pulse. For low transverse
momenta, the contribution from Pair $4(e_1)$ dominates in
$M^{(1)}(\mathbf{p}_1)$, which in Fig.~\ref{Fig5}(a,b) gives rise to the
bright spot at $(p_{1\parallel},p_{1\perp})=(-2\sqrt{U_p},0)$.  Hence, one expects the distribution to be located in the region $-2\sqrt{U_p}\leq p_{2\parallel}\leq 2\sqrt{U_p}$ along the negative $p_{1\parallel}$ half axis. Symmetrization upon electron exchange leads to the occupation of the momentum-space region around the negative $p_{2\parallel}$ half axis, with $-2\sqrt{U_p}\leq p_{1\parallel}\leq 2\sqrt{U_p}$.

For medium transverse momenta the contribution from Pair $3(e_1)$
[corresponding  in  Fig.~\ref{Fig5}(a,b) to the bright ring for
$p_{1\parallel}>0$] becomes comparable to that of Pair $4(e_1)$.
Consequently, the two-electron distribution in Fig.~\ref{Fig7}(e)  spreads
out in the parallel momentum plane. Indeed, the contributions around the positive half axes $p_{n\parallel}>0$ $(n=1,2)$ are now comparable to those along the negative half axes. For large transverse momenta (Fig.~\ref{Fig7}(f)), Pair
$3(e_1)$ dominates over $4(e_1)$, and the distributions moves into the half positive axes. Apart from that, the results show that, the larger the transverse momentum range is, the more concentrated around $p_{n\parallel}=2\sqrt{U_p}$ the electron momentum distributions are.

Up to the present stage, however, causality has not been taken into account.
For instance,  the largest contributions to the partial momentum maps for the
first and second electron [Pair $4(e_1)$ and Orbit $2(e_2)$, respectively],
are not  connected by causality, and therefore their combined contribution
must be discarded. Indeed, the most favorable overall RESI pathway arises
from combining Pair $3(e_1)$ with Orbits $2(e_2)$ (if accessible) and
$3(e_2)$. The large contribution related to Pair $4(e_1)$ is not sufficient
to counter-act the lower tunneling probabilities related to Orbits $4(e_2)$
and $5(e_2)$.  Figs.~\ref{Fig7}(g) to (i) illustrate the consequences.  Since
Pair $3(e_1)$ results in a large yield at $p_{1\parallel}>0$, the electron
momentum distributions are now mostly concentrated along the positive half axes $p_{n\parallel}$ $(n=1,2)$.
This situation persists over all transverse momentum
ranges.

Apart from the reshaping of the distributions, a noteworthy feature in Fig.~\ref{Fig7} is an overall
decrease of a few orders of magnitude in comparison to the situation in
which causality has been disregarded.
Taken altogether, these results show
that causality has a drastic effect on the two-electron momentum
distributions in the RESI pathway.

\section{Conclusions}
\label{Conclusions}

In this work we performed a detailed analysis of the recollision with
subsequent tunneling ionization (RESI) mechanism in laser-induced
nonsequential ionization (NSDI), with emphasis on the implications of
causality. Physically, RESI means that the first electron rescatters with the
core at an instant $t^{\prime}$ and excites a second electron, which tunnel
ionizes at a subsequent time $t$. Causality means that tunnel ionization of
the second electron can only occur after the recollision of the first
electron. Applying the saddle point approximation to the strong-field
expressions, the rescattering and ionization times become complex (since
tunneling does not have a classical counterpart), and the notion of causality
has to be generalized into the complex time plane. We have shown that the
concepts of ``before" and ``after" translate into boundary conditions
limiting the steepest-descent contours to be taken. These boundary conditions
are given by the complex rescattering times of the first electron. Ionization
times of the second electron are casually connected to a rescattering event
if it lies on a steepest descent contour that connects the rescattering time
to the distant future. In practice, this often translates into a simple rule
where only the real parts of the complex times have to be compared. Deviations have been observed, however, for pairs of electron orbits associated with the trailing edge of the pulse, or for momentum regions in which the RESI yield is strongly suppressed.

We illustrated the influence of causality and quantum-interference on
momentum-resolved two-electron distributions. In order to isolated these
effects, we compared distributions for a monochromatic driving field to those
for a few-cycle pulse. While causality does not affect the electron momentum
distributions obtained for monochromatic fields, it significantly affects the
results for a few-cycle pulse. For these pulses, there exist several
competing sets of orbits, whose dominance influences the shapes of the
electron momentum distributions, which generally are highly asymmetric and
concentrated in specific momentum regions.  Causality puts constraints on the
construction of two-electron trajectories, which drastically changes the
momentum regions populated by the distributions.

Causality may have consequences even for monochromatic driving when it is
combined with additional effects, such as bound-state depletion. Depending on
the laser-field intensity and the binding energy of the excited state, such
depletion can be considerable, and if this is the case the ionization time
$t$ lying closest in the future of the rescattering time $t'$ is expected to
contribute most. Such effects go beyond the present work, but may warrant
further investigation.

\section{Acknowledgements}
This work was funded by the UK EPSRC (Grant no EP/D07309X/1) and by the UCL/EPSRC PhD+ scheme. T.S. and C.F.M.F. would like to thank Lancaster University for its kind hospitality.

\end{document}